\documentclass[aip]{revtex4}

\usepackage{graphicx}

\begin{document}

\title{New Electrochemical Characterization Methods for Nanocomposite
Supercapacitor Electrodes}

\author{Jason Ma}
\affiliation{Department of Physics and Astronomy, University of California, Los Angeles,
CA 90024}

\begin{abstract}
Supercapacitor electrodes fabricated from a nanocomposite consisting
of multiwall carbon nanotubes and titanium oxide nanoparticles were characterized electrochemically. 
Conventional electrochemical characterizations cyclic voltammetry and galvanostatic
charge-discharge in a lithium-based electrolyte gave a specific capacitance
of 345 F/g at a current density of 0.1 A/g. New electrochemical
characterization techniques that allow one to obtain the peak capacitance
associated with intercalation and to distinguish between electrostatic
and faradaic charge storage are described and applied to the electrode
measurements. Finally, the maximum energy density obtained was 31 Wh/kg.
\end{abstract}

\maketitle

\section*{I. Introduction }

Supercapacitors, also called ultracapacitors or electrochemical double
layer capacitors (EDLCs), bridge the gap between batteries and conventional
dielectric capacitors \cite{Pandolfo2006,G.A.Snook2011}. The most
common electrode material is carbon-based and include activated carbon
\cite{Pandolfo2006,Frackowiak2001,Zhang2009}, carbon nanotubes \cite{M.Kaempgen2007,C.Liu1999,C.Du2005,M.Kaempgen2009,E.Frackowiak2000,Frackowiak2000,Meyyapp2013},
and graphene \cite{Biswas2010,Y.Wang2009}. To go beyond electrostatic
charge storage in the double layer, pseudocapacitive and faradaic
components (“faradaic components” henceforth) such as conducting polymers
and metal oxides were incorporated into supercapacitors \cite{Frackowiak2001,R.A.Fisher2013,Conway2003,Cericola2012}.
Two ways of incorporating the faradaic material have been employed,
both in an asymmetric (or hybrid) configuration where: (1) a carbon
electrode and a faradaic electrode form the two electrodes of a supercapacitor
\cite{Zheng2003,Simon2008,Q.Wang2006,G.Wang2012,Y.Xue2008}; (2) at
least one of the electrodes is a composite material that combines
advantages of electrochemical double layer charge storage as well
as surface redox or intercalation charge storage \cite{V.Khomenko2006,Z.Chen2009,T.Wang2011,K.Zhang2010,J.Yan2010}. 

Carbon nanotubes (CNT) have a high aspect ratio and form porous networks
with good conductivity. As such, they have been implemented as supercapacitor
electrodes in many instances in the literature \cite{Pandolfo2006,Frackowiak2001,M.Kaempgen2007,C.Du2005,M.Kaempgen2009,E.Frackowiak2000,Frackowiak2000,Meyyapp2013,C.Niu1997,K.H.An2001}.
Titanium oxide is a versatile compound and can be found in several
energy harvesting and energy storage applications, namely photocatalysis,
secondary batteries, and supercapacitors \cite{D.V.Bavykin2006}.
Among the three polymorphs of TiO$_{2}$, the crystal structure of anatase can
accommodate the most Li+ per TiO$_{2}$ unit and is an
attractive energy storage material due to its Li-intercalation capabilities
and cycle life \cite{Z.Yang2009}. TiO$_{2}$ nanowires
have been synthesized by hydrothermal methods,
functioning as the anode of a supercapacitor with carbon nanotubes
acting as the cathode \cite{Q.Wang2006,G.Wang2012}. These are type
(1) asymmetric supercapacitors. 

In this study, the application of a nanocomposite containing TiO$_{2}$
nanoparticles and multiwall carbon nanotubes in a single supercapacitor
electrode is reported. Results from conventional as well as new
electrochemical characterization methods show not only the various
aspects of the energy storage mechanism in the electrode but also the
combination of double layer and faradaic capacitance as a synergy that 
leads to high energy density and good power density in a single electrode.

\section*{II. Experimental Section }

\subsection{Conventional Electrochemical Characterization }

Cyclic voltammetry (CV) as well as galvanostatic cycling (GC) were
performed on the CNT-TiO$_{2}$ composite electrodes. Cyclic
voltammetry of the composite electrode at various scan rates was measured
in 1 M LiClO$_{4}$ in propylene carbonate with lithium
metal sheets as the reference electrode and counter electrode, shown
in Figure 1. The voltage window was 1 V to 2.6 V vs. Li/Li+. A series
of galvanostatic measurements were also performed, two of which are shown
in Figure 2. 

\begin{figure}
\begin{centering}
\includegraphics{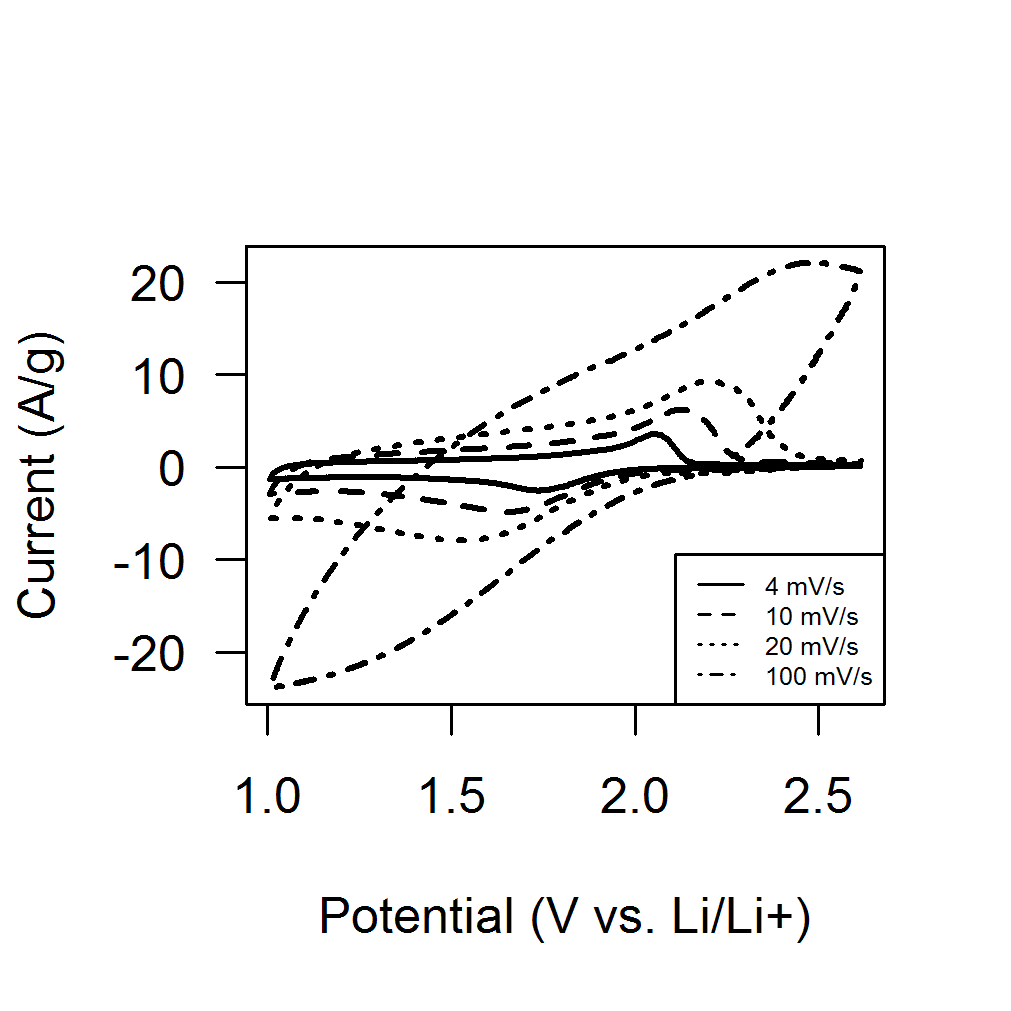}\caption{Cyclic voltammograms (CVs) of CNT-TiO$_{2}$ composite
electrode at various scan rates in 1 M LiClO$_{4}$ in
propylene carbonate with lithium metal sheets as the reference electrode
and counter electrode. }

\par\end{centering}

\end{figure}

\begin{figure}

\begin{centering}
\includegraphics{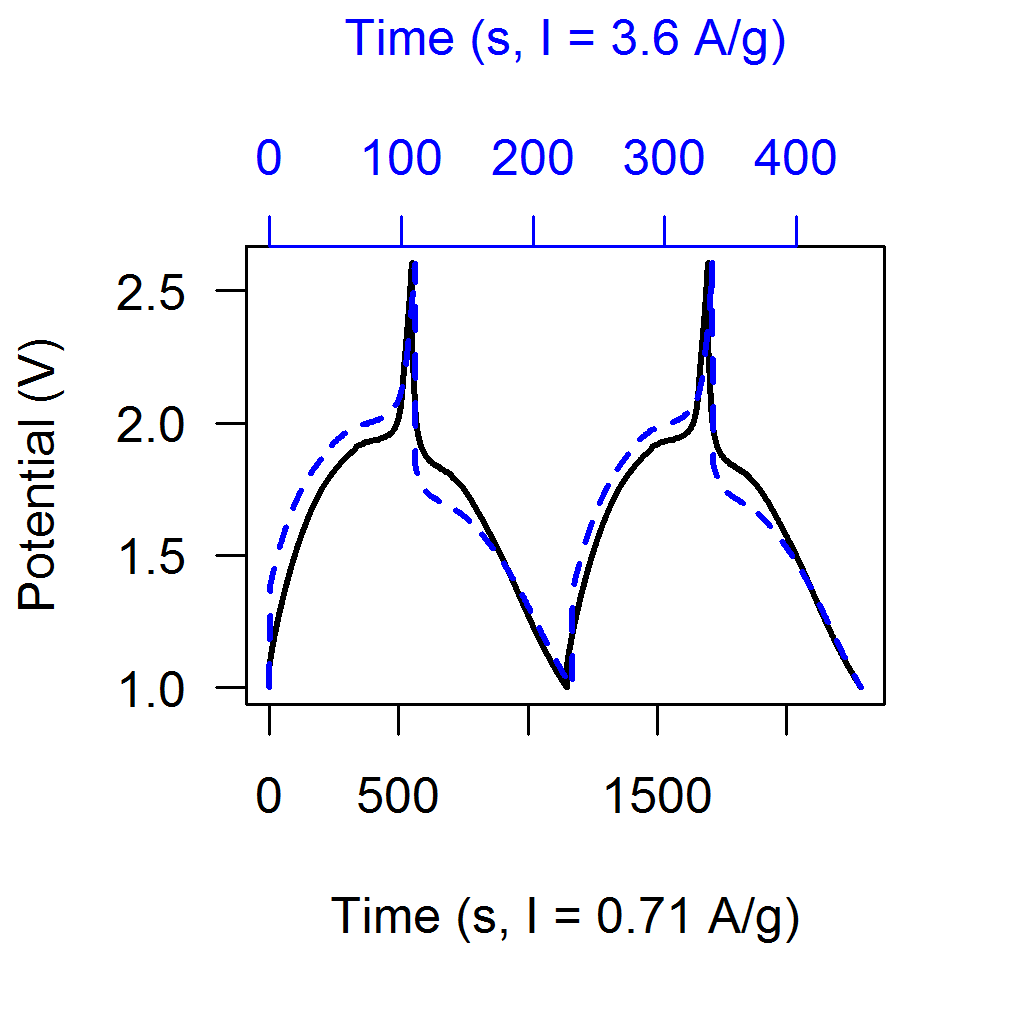}\caption{Galvanostatic measurements of CNT-TiO$_{2}$ composite
electrode with applied currents densities 0.71 A/g and 3.6 A/g.}

\par\end{centering}

\end{figure}
The electrode capacitance is given by $C=Q/\Delta V$ and the specific capacitance by $C_{sp}=C/M$.
The storage capacity of the composite electrode decreases with increasing current but approaches
a constant value, shown in Figure 3 along with the coulombic efficiency.

\begin{figure}
\begin{centering}
\includegraphics{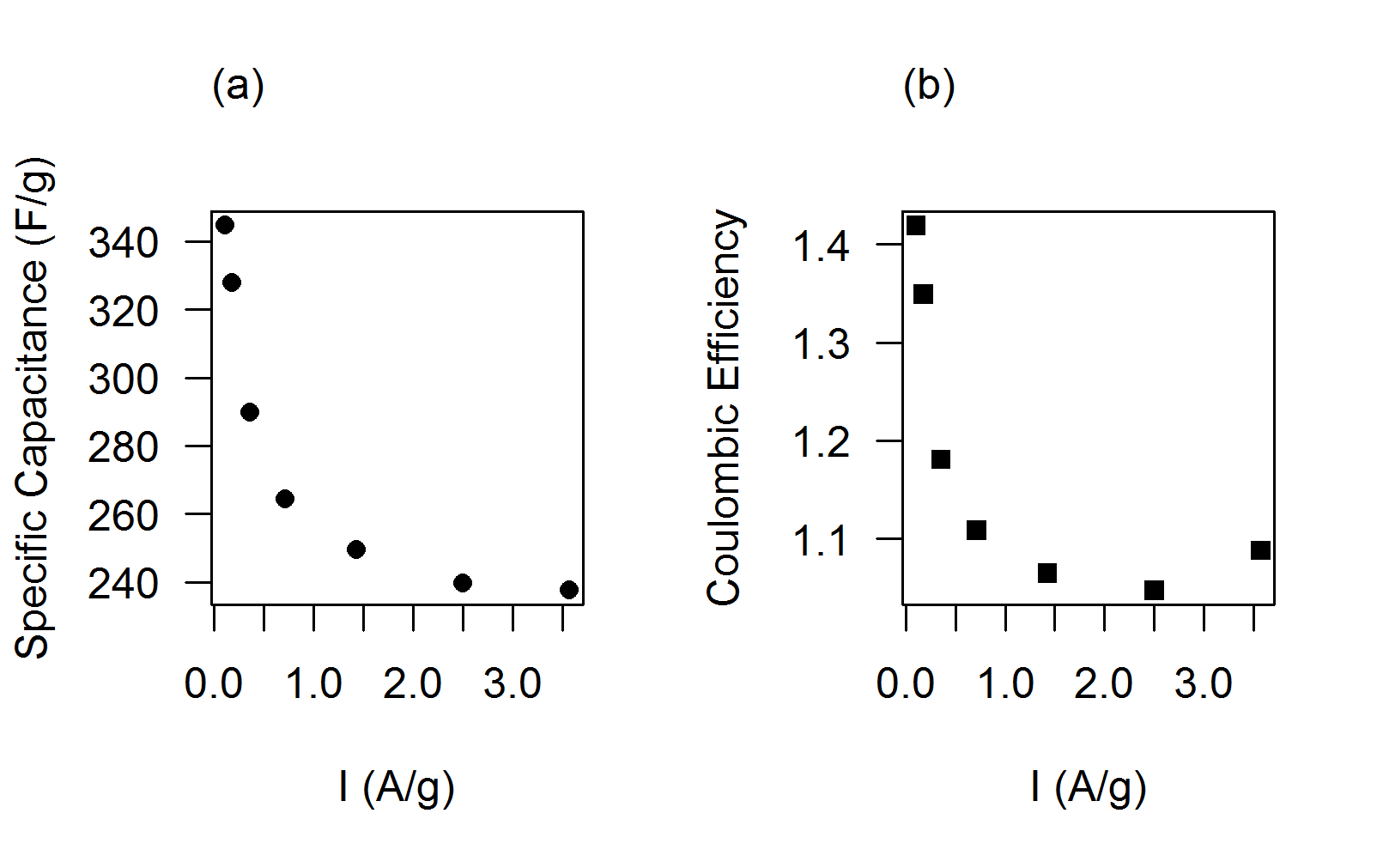}\caption{(a) Specific capacitance vs. current density. (b) Coulombic efficiency.}

\par\end{centering}

\end{figure}

\subsection{New Electrochemical Characterization Methods}

The galvanostatic charging for a given current $I$ is described by
\begin{equation}
V=\frac{I}{C}t+IR_{s}
\end{equation}
where $C$ is the capacitance of the electrode and $R_{s}$ is the
internal resistance. The quantity $IR_{s}$, denoted by $V_{d}$,
is the voltage drop at the beginning of charging and discharging
cycles. The temporal slope $dt/dV$ plotted as a function of the potential
is shown in Figure 4. These plots will be called temporal slope voltammograms
(TSVs) in this article. Similar to CVs, the anodic peaks and cathodic peaks shift as the applied current increases.
However, whereas the faradaic/intercalation peaks flatten as the
scan rates increase in CVs, the temporal slope peaks remain sharp
even for large current densities. 

\begin{figure}

\begin{centering}
\includegraphics{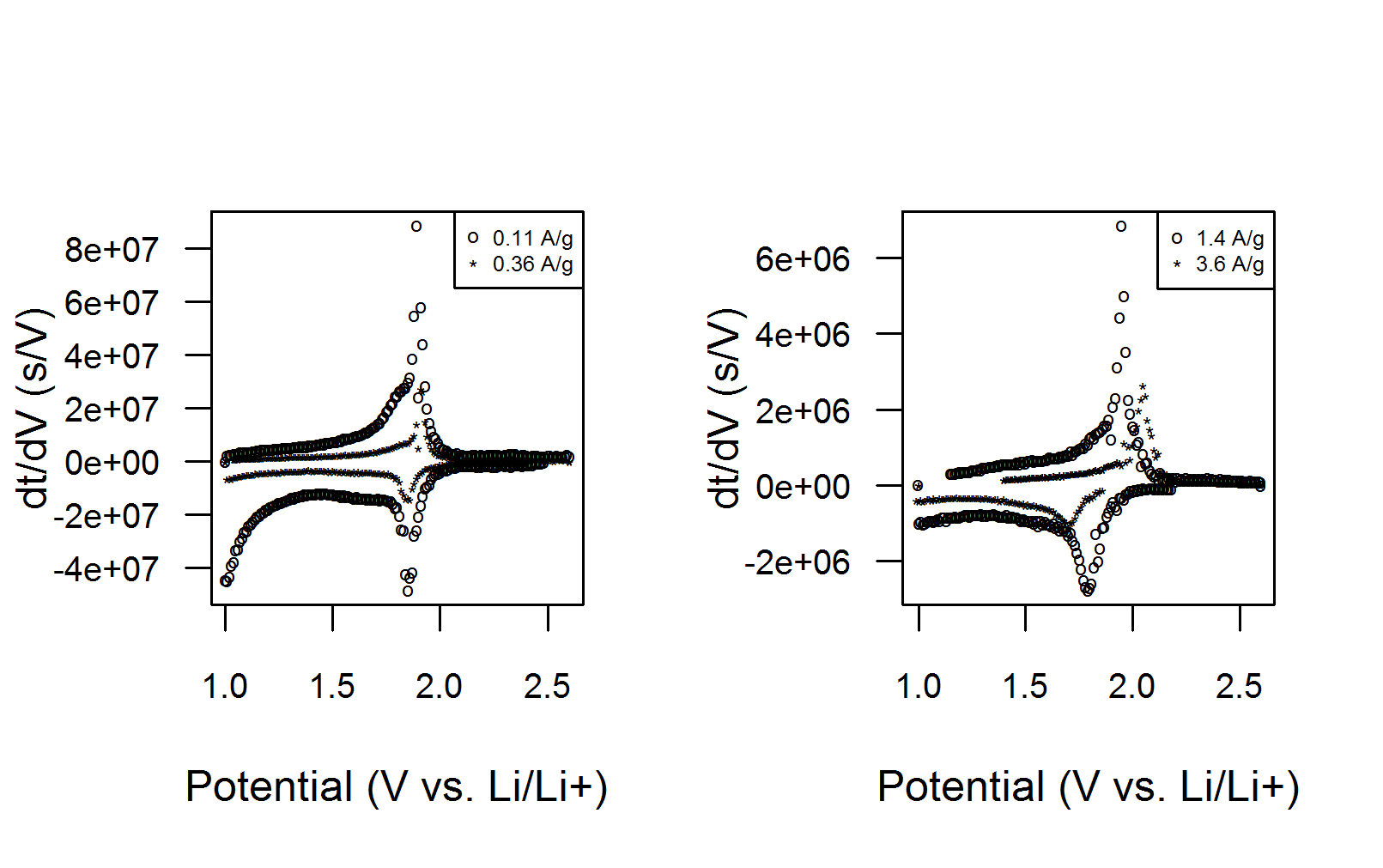}\caption{Temporal slope voltammograms were constructed from galvanostatic measurements at current densities 0.11, 0.36, 1.4, and 3.6 A/g.}

\par\end{centering}

\end{figure}

As shown in Figure 5a, as $V_{d}$ vanishes, both anodic and cathodic
peak potentials approach the same value, the standard potential for
the intercalation reaction. A plot of the
temporal slope peaks versus $I^{-1}$ shows a linear relationship
(Figure 5b). Linear regression gives the peak capacitances at $V_{p}$,
2410 F/g for Li extraction and 1450 F/g for Li insertion. As a point
of reference, the theoretical specific capacitance of TiO$_{2}$
assuming one Li+ ion per TiO$_{2}$ molecule is 2295 F/g,
calculated using the formula $FV_{p}/\rho_{TiO_{2}}$, where $F$
is the faraday constant and $\rho_{TiO_{2}}$ is the TiO$_{2}$
molar mass. The peak capacitance at the standard potential of the
faradaic reaction could be an important figure of merit in characterizing
and optimizing nanocomposites as the mass normalized quantity is
nearly independent of the double layer contributions. 

In addition to temporal slope voltammograms, a new method to
distinguish between electrostatic and faradaic contributions to the 
energy storage capacity warrants discussion. The
distinction is important because one is often interested in the effect
of nanoparticle size on the faradaic reactions and whether the faradaic
process involves only surface sites or the bulk \cite{J.Wang2007}.
The differential form of the capacitor equation

\begin{equation}
dq=CdV+VdC
\end{equation}
includes both capacitor behavior, where the capacitance is a constant that depends only on 
the capacitor\textsc{\char13}s geometry and dielectric material, and battery behavior, where the potential
is ideally constant and the storage capacity varies linearly with the state of charge.
In the case where the potential is the experimentally tunable parameter, charge conservation requires 

\begin{equation}
\frac{dq}{dV}=C+V\frac{dC}{dV}
\end{equation}

The current associated with changes
in the double layer is given by $I_{dl}=C_{dl}dV/dt$ and
the current due to reversible charge transfer reactions
at the electrode surface is described by \cite{Bard2000}

\begin{equation}
I_{F}=\frac{(nF)^{2}}{RT}\delta AC_{O}^{o}\frac{e^{\mathcal{F}(V^{o}-V)}}{\left(1+e^{\mathcal{F}(V^{o}-V)}\right)^{2}}\frac{dV}{dt}
\end{equation}
where $\mathcal{F}=nF/RT$, $C_{O}^{o}$ is the starting ion concentration at the electrode-electrolyte interface,
$\delta$ is the width of the diffusion layer, $A$ is the area of
the TiO$_{2}$ nanoparticles, and $V^{o}$ is the standard
potential of the reaction. It follows that the charge stored is 

\begin{equation}
q(V)=nF\delta AC_{O}^{o}\left(\frac{1}{1+e^{\mathcal{F}(V^{o}-V)}}-\frac{1}{1+e^{\mathcal{F}V^{o}}}\right)+C_{dl}V
\end{equation}

The charge accumulation as a function of potential can be obtained
from galvanostatic measurements, as shown in Figure 6.
The double layer capacitance can be obtained from the slope of $q(V)$
for sufficiently large potentials, after available TiO$_{2}$
lattice sites are fully intercalated. 

\begin{figure}
\begin{centering}
\includegraphics{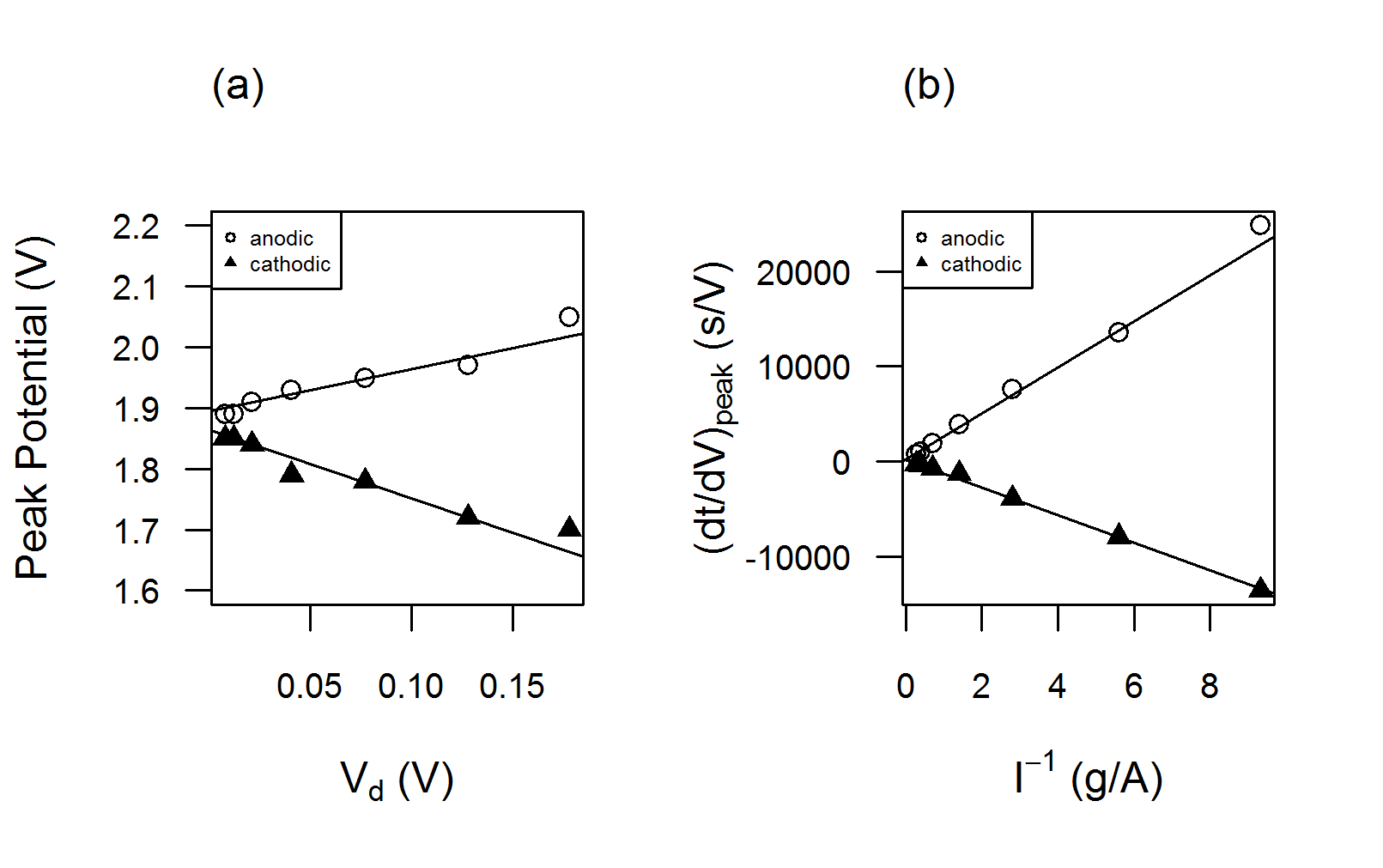}
\par\end{centering}

\caption{(a) The peak potential is linear in the voltage drop $V_{d}$. The
anodic and cathodic intercepts are both 1.9 V. (b) Temporal slope
peaks are linear in $I^{-1}$. The absolute values of the two linear
regression slopes yield peak capacitances.}

\end{figure}

\begin{figure}
\begin{centering}
\includegraphics{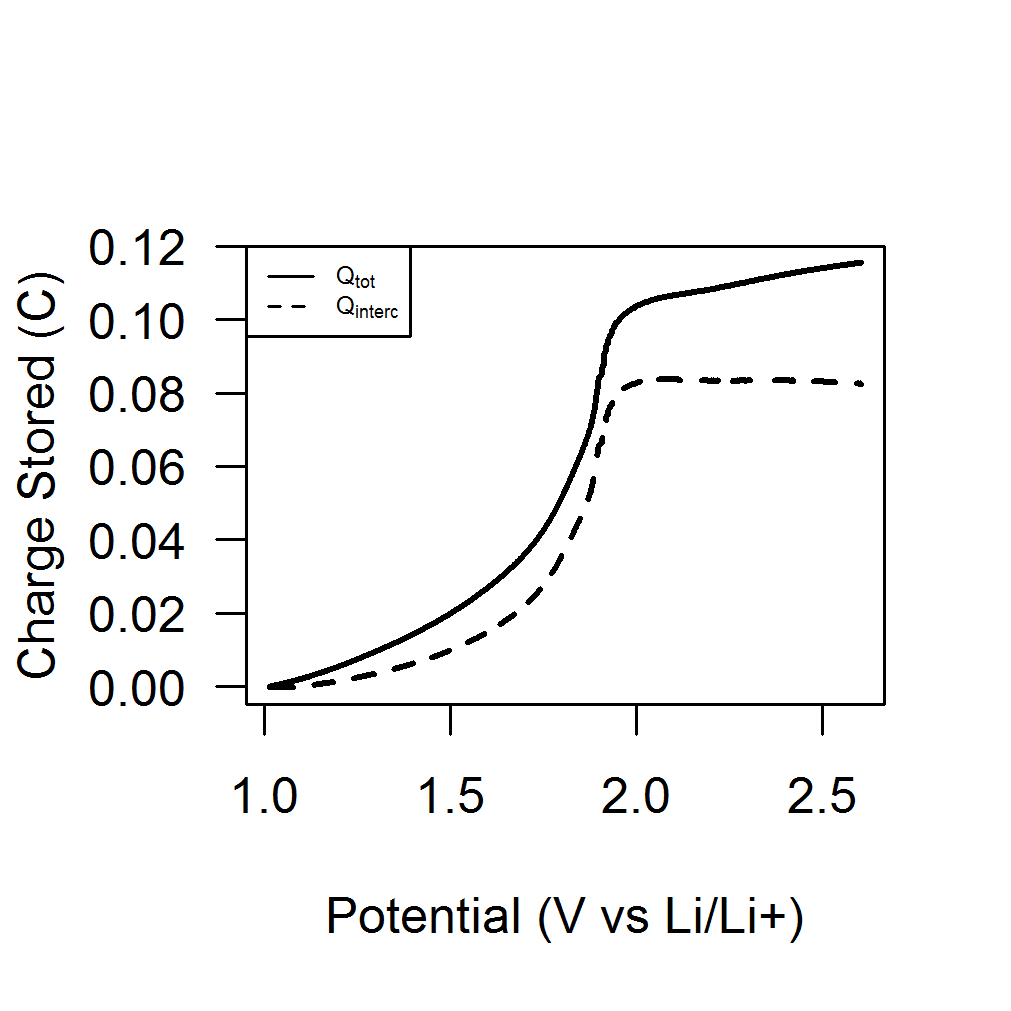}
\par\end{centering}

\caption{Charge accumulated as a function of potential for a current density
of 0.1 A/g.}
\end{figure}

\section*{III. Results and Discussion }

Dielectric capacitors experience changes in potential at a constant storage capacity,
whereas batteries experience changes in the storage capacity at
a constant potential. The methods outlined in this article enable researchers
to electrochemically characterize nanocomposites that exhibit both capacitor-like
and battery-like behaviors.

The performance of the CNT-TiO$_{2}$ electrode can be
compared to lithium-ion batteries and other electrochemical capacitors
in the form of a ragone chart, shown in Figure 7. The storage capacity
and the voltage range can be converted to energy density and power
density with the formulas 

\[
ED=\frac{1}{8}\frac{C_{sp}\text{\ensuremath{\Delta}}V^{2}}{3.6},\quad PD=\frac{\text{\ensuremath{\Delta}}V^{2}}{8MR_{s}}
\]

The synergy between the two materials gives rise to high energy and
good power. From the perspective of the high energy density Li-intercalation
material, namely the TiO$_{2}$ nanoparticles, the addition
of carbon nanotubes led to a significant increase in power density. 

\begin{figure}
\begin{centering}
\includegraphics{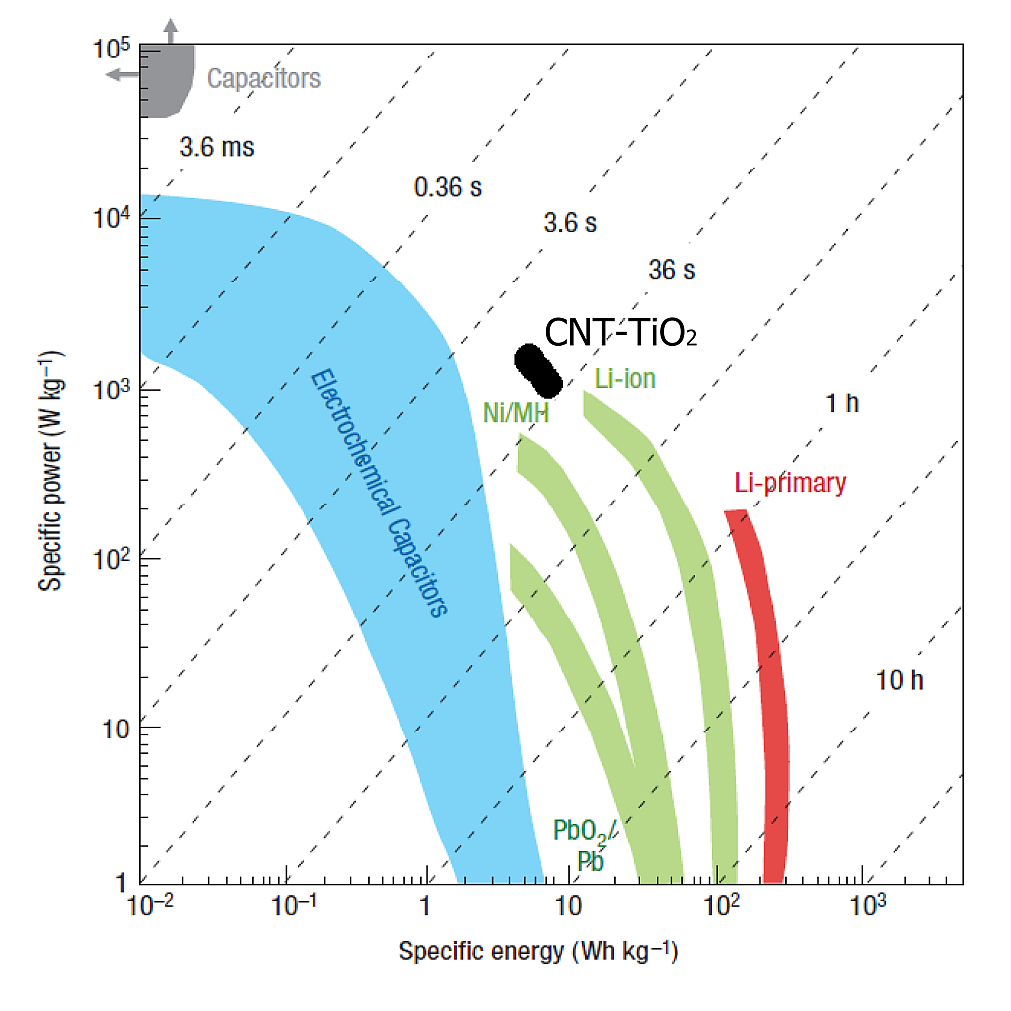}
\par\end{centering}

\caption{Ragone chart showing the performance of the CNT-TiO$_{2}$
electrode with respect to Li-ion batteries and other electrochemical
capacitors. {[}Adapted from Ref. 18 with permission.{]}}

\end{figure}

\section*{IV. Conclusion}

A supercapacitor electrode was fabricated from a nanocomposite consisting
of multiwall carbon nanotubes and titanium oxide nanoparticles. Conventional
electrochemical characterizations cyclic voltammetry and galvanostatic
cycling gave a specific capacitance of 345 F/g at a current density
of 0.1 A/g. New electrochemical characterization techniques derived from galvanostatic measurements
allow one to obtain the peak capacitance associated with intercalation
and to distinguish between electrostatic and faradaic contributions to the total charge stored. 
The new techniques show that most of the charge is stored faradaically, via the intercalation
mechanism. The double layer charge storage mainly attributed to carbon
nanotubes brought significant improvement in power density to the
faradaic material. As a nanocomposite, CNT-TiO$_{2}$
achieved a maximum energy density of 31 Wh/kg.

\section*{Acknowledgement}

Helpful discussions from George Gr\"{u}ner, Bruce Dunn, Ryan Maloney,
Veronica Augustyn as well as measurement support and TiO$_{2}$
nanoparticles from Veronica and Jesse Ko are acknowledged and appreciated.

\end{document}